\documentclass[prb,twocolumn,showpacs]{revtex4}
\usepackage{graphicx}
\usepackage{amsmath}
\usepackage{amssymb}

\newcommand{\sss}{\scriptstyle}
\def\lsim{\
  \lower-1.2pt\vbox{\hbox{\rlap{$<$}\lower5pt\vbox{\hbox{$\sim$}}}}\ }
\def\gsim{\
  \lower-1.2pt\vbox{\hbox{\rlap{$>$}\lower5pt\vbox{\hbox{$\sim$}}}}\ }

\begin{document}


\title{Electromagnetic and phonon modes
    for superfluid $He^4$ with a disk resonator }

\author{V.M.~Loktev}
\email{vloktev@bitp.kiev.ua}
\author{M.D.~Tomchenko}
\email{mtomchenko@bitp.kiev.ua}
\affiliation{Bogolyubov Institute for Theoretical Physics,
        14-b Metrologicheskaya Street, Kiev, 03680, Ukraine}

\date{\today}

\begin{abstract}
We  find the distribution of the electromagnetic field inside and outside
    a dielectric disk resonator placed in He-II. It is shown that
    this field consists of a collection of  ``circular''  (c-) photons. The
    wave function $\Psi_{c}$ of a
    c-phonon for the He-II + disk system is
    calculated in the zero-order  approximation in interaction. Due to the symmetry of the problem, the
    structure of $\Psi_{c}$ is such that a c-phonon possesses,
    similarly to a c-photon of the resonator, a definite energy  and an angular momentum with
        respect to the disk axis, but it does not possess a definite
        momentum in the disk plane.
\end{abstract}

\maketitle

   \section{Inroduction}
 \noindent   In the recent years, some interesting
    and, in a certain sense, unexpected effects were discovered in the  experiments \cite{svh2,svh3}. Namely,
    a supernarrow absorption line at the frequency of the roton minimum  was registered in the spectrum of a
dielectric disk SHF resonator placed in liquid He-4. In an external constant electric field,
    the line is split into two ones. At the switching-on of a heat gun directed along a tangent to the disk, the
    absorption line is transformed in an emission line. These effects have no explanation yet, though the line itself is related, undoubtedly, to
     a high density of states of He-II at energies close to the roton one \cite{svh3}.

     To explain the origin of the line and its specific features, it is necessary to determine, first of all, the electromagnetic (EM) field of a
     resonator, as well as the wave functions of a phonon and a roton for helium with an immersed disk. The present work is devoted to this problem.

 \section{Electromagnetic field of a disk resonator}

         In the experiments described in \cite{svh2,svh3}, a variable inhomogeneous
           field with the strength $E\leq 10^4\,\mbox{V}/\mbox{m}$ was induced in a resonator. The field was mainly concentrated in a disk and created
           the deformations of a resonator which are pulsating in time and space.
           However, at the attained values of $E,$ the total deformation of a disk was very small
           --- at most $0.1 \mbox{\AA}$ \cite{smag} for a quartz resonator.
           Similar weak pulsations should play no
            role in the phenomena under study. Therefore, it is obvious that a roton is excited
           by the SHF field of a circular EM wave pulsating on the rim of a disk, rather than by deformations of the disk.
           In what follow, we will calculate the EM field of a resonator.

     In the experiments, the sizes of disk resonators were
      approximately identical. In \cite{svh2} and \cite{svh3}, the resonators were fabricated of quartz
      and leucosapphire, respectively. The results obtained for the shape and the width of a roton
     line are close, but the numbers of the azimuth mode (for the roton frequency) are different.
     Below, we will obtain the general formulas for the EM field of a resonator and analyze the numerical values for the experimental conditions in \cite{svh2}.
     Let us consider the EM wave propagating in a quartz resonator with the shape of a disk
     with the thickness $h_d=1\,\mbox{mm}$ and the radius $R_d=9.5\,\mbox{mm}$.
     The dielectric permittivity tensor $\varepsilon_{jk}$ for the quartz under study is diagonal in the coordinate system (CS),
     whose $Z$ axis coincides with the geometric axis (it is also optical)
     of a resonator; in this case,  $\varepsilon _z  = 4.63$, and $\varepsilon _{\bot} =
      4.43$ in perpendicular directions \cite{d}.

       In calculations of the EM field, we are based on the Maxwell equations
       in a medium:
          \begin{equation}
          div \textbf{D}=0, \quad div \textbf{B}=0,
       \label{2-1} \end{equation}
       \begin{equation}
         rot \textbf{E}=-\frac{1}{c}\frac{\partial \textbf{B}}{\partial t}, \ \
         rot \textbf{H}=\frac{1}{c}\frac{\partial \textbf{D}}{\partial t},
       \label{2-2} \end{equation}
        \begin{equation}
           \textbf{D}=\hat{\varepsilon}\textbf{E},
       \label{2-3} \end{equation}
       where $c$ is the light velocity in vacuum.
      For quartz and helium, $\mu \approx 1$, therefore,
      $\textbf{B}=\textbf{H}$.

       We now find the
      vector potential $\textbf{A}$ connected with $\textbf{E}$ and $\textbf{H}$ by the relations
       \begin{equation}
         \textbf{E}=-\frac{1}{c}\frac{\partial \textbf{A}}{\partial t} - \nabla\varphi_{el},
               \label{2-4} \end{equation}
        \begin{equation}
         \textbf{B}= rot \textbf{A}.
               \label{2-5} \end{equation}
      We use the transverse calibration $\varphi_{el}=0$ and
      pass into a cylindrical CS (CCS) $\rho, \varphi, z$ with the origin at the disk center and the
     $Z$ axis coinciding with the axis of a resonator. In the CCS, the tensor
     $\varepsilon_{jk}$ is diagonal: $\varepsilon _{xx} = \varepsilon _{yy} =\varepsilon _{\bot}$,
     $\varepsilon _{zz} =\varepsilon _{z}$. For the field in quartz, relations (\ref{2-1}) and (\ref{2-4}) yield
     \begin{equation}
         div\textbf{A}(\textbf{r},t)=\left (1-\frac{\varepsilon_{z}}{\varepsilon_{\bot}}\right )\frac{\partial A_{z}}{\partial z}
         +  f_{d}(\textbf{r}),
               \label{2-6} \end{equation}
       where $f_{d}$ is some function independent of $t$. Since we are interested in EM waves, we can take $f_{d}=0$. With the help of (\ref{2-2})
       and (\ref{2-5}), we obtain the following equation for $\textbf{A}$:
        \begin{equation}
        \triangle \textbf{A} - \frac{\hat{\varepsilon}}{c^2}\frac{\partial^{2} \textbf{A}}{\partial t^{2}}
         + \left (\frac{\varepsilon_{z}}{\varepsilon_{\bot}}-1\right )\nabla\frac{\partial A_{z}}{\partial z}
         =0.
               \label{2-7} \end{equation}
       For quartz, the values of $\varepsilon_{z}$ and
       $\varepsilon_{\bot}$ are close. Therefore, we can neglect their
       difference and consider that $\varepsilon_{z}=\varepsilon_{\bot}=\varepsilon_{d}\approx 4.63$, which simplifies the equation:
        \begin{equation}
        \triangle \textbf{A} - \frac{1}{c_{\perp}^2}\frac{\partial^{2} \textbf{A}}{\partial t^{2}}
                =0,  \quad c_{\perp} = \frac{c}{\sqrt{\varepsilon_{\perp}}}.
               \label{2-8} \end{equation}
                 This equation has a solution $\textbf{A}$
         directed identically at all points of a resonator and another
         solution directed according to the symmetry of the disk with the $\rho$-,
         $\varphi$-, and $z$-components. It is natural to expect that a
         resonator amplifies maximally those components of the field which correspond to its symmetry.
         It follows from the experiment \cite{d} that this is true,
         and, in addition, the principal components of the field $\textbf{E}$ near
          a resonator are the $\rho$- and $\varphi$-components, whereas
          the value of the $z$-component is less by three orders. Therefore, we neglect the latter
          and consider that the field $\textbf{A}$ in a resonator and in helium has only $\rho$- and
          $\varphi$-components.

      The equation for the field outside a resonator (in helium)
      has the form
        \begin{equation}
  \triangle \textbf{A} - \frac{1}{c_{h}^2}\frac{\partial^{2} \textbf{A}}{\partial t^{2}} =0,
     \label{2b} \end{equation}
      where $c_{h}=c/\sqrt{\varepsilon_{h}}$,
      $\varepsilon_{h}=1.057$ (here and below, $h$ and $d$ mean, respectively, helium and a disk).
           In order to determine $\textbf{A},$ it is necessary to solve Eqs. (\ref{2-8}) and
      (\ref{2b}) with regard for boundary conditions (BCs) on the surface
      of a resonator:
     \begin{equation}
     E^{d}_{\parallel}= E^{h}_{\parallel}, \ \ H^{d}_{\parallel}= H^{h}_{\parallel}   \label{gr1} \end{equation}
     and, if there are no extrinsic charges,
      \begin{equation}
     B^{d}_{\perp}= B^{h}_{\perp}, \ \ D^{d}_{\perp}= D^{h}_{\perp}   \label{gr2} \end{equation}
      (here, the symbols $\parallel$ and $\perp$ indicate the relations to the surface, whereas the symbol $\perp$
      in the other cases means the relation to the disk axis (the $Z$ axis)).

       We now calculate the field $\textbf{A}$ outside and inside a disk. The general form of a solution $\textbf{A}(\rho, \varphi, z)$ is unknown else
       and, generally speaking, complicated.
        In principle, the field can depend on the shapes and the sizes of a container and antennas \cite{d}
       (for example, in the experiments in \cite{svh2,svh3}, two antennas are positioned in the disk plane on two sides from it at a distance of $13\,\mbox{mm}, i.e., \approx 1.37 R_{d}$ from the disk axis).
        The geometry of a resonator is such that the field inside a disk
        can be determined with the use of the separation of variables:
                 \begin{eqnarray}
 && \textbf{A}(\rho ,\varphi,z,t) =  e^{-i\omega t}\left [\int\limits_{-\infty}^{\infty}dQ_{z}
  \textbf{a}_{Q_z}(\rho ,\varphi) \right. \nonumber \\
  && \left. \times[b_{c}(Q_z)\cos{(Q_z z)}
      + b_{s}(Q_z)\sin{(Q_z z)}] \right ] + \mbox{c.c.}
         \label{2} \end{eqnarray}
        Here, we took into account that the observed field is real. The solution contains no sines, because the system is symmetric relative to the reflection
         $z \rightarrow -z$.  It is known from the experiment that a disk enhances the field
        $\textbf{E}$ mainly inside itself. Outside the disk, the field is slight, rapidly decreases,
         and disappears practically at a distance of ~2 mm from the disk.
        Therefore, we assume that the structure of the solution outside the disk is such that we can approximately separate the variables $z$ and (on the other hand)
        $\rho$, $\varphi$ according to (\ref{2}).

   Since the $z$-component $\textbf{A}$ is small, we can write  $\textbf{a}_{Q_z}(\rho ,\varphi)=
    a^{\rho}_{Q_z}(\rho ,\varphi) \textbf{e}_{\rho}+a^{\varphi}_{Q_z}(\rho ,\varphi) \textbf{e}_{\varphi}$.
   Then relation (\ref{2-8}) yields the equation for $\textbf{a}_{Q_z}$:
  \begin{equation}
  \triangle_{\rho,\varphi} \textbf{a}_{Q_z} + (Q^{d}_{\rho})^{2} \textbf{a}_{Q_z}  =  0,
     \label{3-1} \end{equation}
     where $Q^{d}_{\rho}$ depends on $Q_{z}$:
     \begin{equation}
  Q^{d}_{\rho} = \sqrt{(Q^{d})^2- Q_{z}^2 },  \ \
     Q^{d} = \omega/c_{\perp} = 2\pi\nu/c_{\perp}.
     \label{3-2} \end{equation}
     After simple calculations, we get the general solution of Eq.
     (\ref{3-1}) for a real $ Q^{d}_{\rho}$:
     \begin{eqnarray}
  &\textbf{a}_{Q_z}(\rho ,\varphi) &= \sum\limits_{l} \int dQ_{z}g_{l}(Q_{z})e^{il\varphi} \label{4-1}  \\
   && \times\left \{[J_{l-1} (Q^{d}_{\rho}\rho)-c_{l}(Q_{z})J_{l+1} (Q^{d}_{\rho}\rho)]\textbf{e}_{\rho}
   \right. \nonumber \\ &&+ \left.
   i\textbf{e}_{\varphi}[J_{l-1} (Q^{d}_{\rho}\rho)+c_{l}(Q_{z})J_{l+1} (Q^{d}_{\rho}\rho)]\right \}.
     \nonumber \end{eqnarray}
     Here, $l$ is an integer, $J_{l} (x)$ is the
     Bessel function, and $g_{l}(Q_{z})$ and $c_{l}(Q_{z})$ are constants. The second independent solution of Eq. (\ref{3-1}) proportional to the Neumann functions
     $N_{l\pm 1} (Q^{d}_{\rho}\rho)$ is omitted, because it tends to infinity as $\rho \rightarrow 0$.
     The radial wave number $Q^{d}_{\rho}$ in (\ref{4-1}) is determined, according to (\ref{3-2}), by the value of $Q_z$; $Q^{d}_{\rho}$ is positive at
       $Q_z < Q^{d}$ and imaginary at $Q_z > Q^{d}$. For the imaginary argument, we have $J_l (ix)=i^l I_l(x)$ \cite{tih}.
       The plot of the function $J_{l}(x)$ for $l=67$
      is given in Fig. 1.   The function $J_{l}(x)$ oscillates outside the disk, whereas $I_l(x)$
      increases monotonously and rapidly for all $x$. The EM pumping field creates some field $\textbf{A}$ with a given frequency
      $\nu$ in the disk and outside it,
      and this field increases in a resonance manner at definite values of $\nu$, $Q_z,$ and $l$. We do not calculate the exact condition for a resonance and the width of the resonant $l$-mode,
           because it is easy to establish which modes of (\ref{4-1}) are observed with the help of experimental data. It can be expected that
      the approximate condition for a resonance consists in the proximity of the EM field on the surface of a resonator to zero (see (\ref{gal})).
       \begin{figure}[h]
\centerline{\includegraphics[width=85mm]{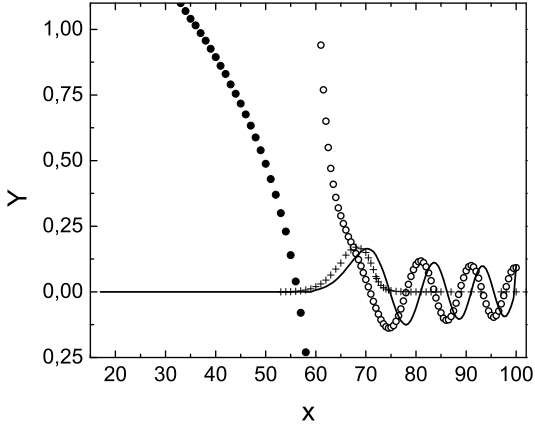}}
\caption{ \ Solid line --- the Bessel function $J_{l_{rot}+1}(x)=J_{67}(x)$;
 $\circ\circ\circ$ --- the Neumann function $-N_{67}(x)$ which grows very rapidly at $x\lsim 60$. Therefore,
  the function  $lg(lg(-N_{67}(x)))$ denoted by $\bullet\bullet\bullet$ is shown in this region.
  $+++$ --- the functions  $\sqrt{J^2_{65}(x)+c^2_{66}
      J^2_{67}(x)}$ at $x\leq x_{0}$ and $J_{65}(x_{0})\sqrt{n^2_{65}(xQ^{h}_{1}/Q_{1})+b^2_{66}
      n^2_{67}(xQ^{h}_{1}/Q_{1})}$ at $x > x_{0}$ present the radial behavior of the field $A$ for $z=0$
  (in this case, $x_{0}=Q_{1}R_{d} \approx  72.1$ --- disk edge, $Q^{h}_{1}/Q_{1}\approx 0.34$).}
\end{figure}

  Experiments revealed various resonance modes, from which the first
        $z$- and the first radial harmonics were studied in detail.
         By $\mu_{l}^{(n_{\rho})}$ ($n_{\rho}= 1, 2, 3, \ldots$), we denote values of $x$, for which $J_l (x)=0$.
            The first radial harmonic ($n_{\rho}=1$) means that the field in the disk is distributed
        over the radius so that it includes only the first half-wave of the function $J_{l}(x)$ and becomes almost zero near the disk edge. Therefore,
        $Q^{d}_{\rho}R_{d} \approx  \mu_{l}^{(1)}$. To be more exact, as $\rho$ increases from zero to $R_{d},$ the field $|\textbf{A}|$ increases firstly, attains a maximum near the disk edge ($\rho\approx R_{d}-0.5\,\mbox{mm}$),
     then decreases, and takes $\sim 1/8$ of the maximum value at the disk edge ($\rho = R_{d}$).  The first $z$-harmonic means that
        the distribution of the field over $z$ is close to $\cos{(z\pi/\tilde{h}_d)}$.
                                   On the upper and lower surfaces of the disk $z=\pm h_{d}/2$, the field
            $A$ is less approximately by 8 times than that in the middle plane of the disk
            ($z=0$) at the same $\rho$ and $\varphi$, which yields $\tilde{h}_d \approx 1.087 h_{d}$ and
            $Q_{z} = \pi/\tilde{h}_d$. This allows us to uniquely determine $Q^{d}_{\rho}$:
             $Q_{\rho}^{d}\equiv Q_{1} \approx 72.143/R_{d}$.
             The BCs (\ref{gr1}) and (\ref{gr2})
       yield $A_{\rho}(\rho=R_{d})=0$, which gives
        \begin{equation}
       c_{l}(Q_{z}) = J_{l-1} (Q^{d}_{\rho}R_{d})/J_{l+1}(Q^{d}_{\rho}R_{d}).
       \label{c-l2} \end{equation}

         The roton line is
     observed for a certain mode characterized by $l_{rot}$. In \cite{svh2}, the quantity $l_{rot}$ was defined as the ratio of the frequency to the step between modes, and its value was estimated as
      $l_{rot} \approx 78$.  However, the approximate condition of resonance (\ref{gal}) implies that the connection between the frequency $\nu$ and $l$ is not strictly linear,
        and the step must somewhat increase with decrease in $\nu$.
           By averaging $|\textbf{A}|$ over time, we obtain $|\textbf{A}| \sim f(Q_{1}\rho)=\sqrt{J^2_{l-1}(Q_{1}\rho)+c^2_{l}
      J^2_{l+1}(Q_{1}\rho)}$. According to experiments, the maximum value $f_{m}$ of the function $f$ on the interval $\rho = 0\div R_{d}$ is attained at $\rho_{m}\approx R_{d}-0.5\,\mbox{mm}$. For $\tilde{h}_d$ obtained above, such a value of $\rho_{m}$ follows from calculations for $l_{rot} \approx 66$. In this case,  $f_{m}\approx 0.17$, and
       $f\approx 0.05$ at the disk edge. The experiment gives that $A$ at the edge is equal to $1/6 \div 1/10$ of the maximum value at the same height, i.e. $f\approx 0.02\div 0.03$.

       We now obtain the final
       solution for the field $\textbf{A}$ inside the disk:
                \begin{eqnarray}
      \textbf{A}_{d}(\rho ,\varphi,z,t) &=&  A_{m}e^{il\varphi -i\omega t}
       \cos{(\pi z/\tilde{h}_d )}\nonumber \\
       &\times & \left \{[J_{l-1} (Q_{1}\rho)-c_{l}J_{l+1} (Q_{1}\rho)]\textbf{e}_{\rho}
      \right.  \label{Ain1}  \\ &+& \left.
   i\textbf{e}_{\varphi}[J_{l-1} (Q_{1}\rho)+c_{l}J_{l+1} (Q_{1}\rho)]\right \} + \mbox{c.c.},
         \nonumber \end{eqnarray}
           \begin{eqnarray}
        A_{m} &=& \frac{E_{m}c}{2\omega f_{m}} \approx \frac{2.94E_{m}c}{\omega}, \quad
        Q_{1} \approx 72.143/R_{d}, \nonumber \\
       l&=&66,  \quad c_{l}\approx 0.265, \quad \tilde{h}_d \approx  1.087 h_{d}.
         \label{Ain2} \end{eqnarray}

      The experiment
      \cite{svh2} indicates that two counter circular EM waves propagate on the disk, and the amplitude of one of the waves is larger by
      2 orders than that of the second one. Below, we will neglect the weaker wave characterized by a different sign of $l$.

        It is worth to note that the circular EM wave (\ref{Ain1})
        has no definite $z$-momentum. Indeed, acting by the operator
      $\hat{P}_z=-i\hbar\partial/\partial z$ on state (\ref{Ain1}), we obtain $\sim \sin{(\pi z/\tilde{h}_d )}$,
      i.e. the state is changed.

      Thus,
       the system of waves in a resonator is characterized by three quantum
       numbers:  $l$,   $n_{\rho},$ and $Q_z$.
  For the field in the disk, we possess solution (\ref{Ain1}), (\ref{Ain2}) with
        $Q_z =\pi/\tilde{h}_d$, $n_{\rho}=1,$ and different $l$.
           Since the field is insignificant near the disk edge, the resonance frequencies $\nu (l,n_{\rho},Q_{z})$
        are determined approximately by the equation $A(\rho=R_{d})=0$. Whence we get
         $J_{l-1}
       (Q^{d}_{\rho}R_{d})=0$, i.e.
       $Q^{d}_{\rho}R_{d} = \mu_{l-1}^{(n_{\rho})},$ or
       \begin{equation}
      \sqrt{\left (\frac{2\pi\nu R_{d}}{c}\right )^2 \varepsilon_{\perp} - (Q_z R_{d} )^2} = \mu_{l-1}^{(n_{\rho})}.
     \label{gal} \end{equation}
     This is an approximate condition of resonance.
        For each mode $\nu(l,n_{\rho},Q_z)$ (\ref{gal}), the distribution of the field $\textbf{A}$ in the disk at large $l$ is similar to a circular gallery.
       Such resonance modes are called ``whispering-gallery modes'',
        because it was noted else in ancient times that a word pronounced by whisper at certain places of a
         circular gallery at large temples is heard at a remote part of a temple.

          We are interesting in the modes $\nu(l,1,\pi/\tilde{h}_d)$. At $T$=1.4\,K, the roton absorption line was
         observed at the frequency $\nu_{rot}=180.3\,\mbox{GHz}$ which corresponds, as shown above, to $l=l_{rot}\approx  66$. We consider that, at $l\gg 1,$
        the following relation is true \cite{y}:
        \begin{equation}
        \mu_{l}^{(1)} \approx l+1.856\,l^{1/3} + 1.033\,l^{-1/3}.
            \label{x0} \end{equation}
         This yields $\mu_{l_{rot}-1}^{(1)}=\mu_{65}^{(1)} \approx 72.719$.
         Let us denote the harmonic $\nu(l_{rot},1,\pi/\tilde{h}_d)$ by $\nu_{l_{rot}}$. For the sizes $R_d$ and $h_d$ taken from \cite{svh2},
        relation (\ref{gal}) yields $\nu_{l_{rot}} \approx 1.007 \nu_{rot}=\nu_{rot} +1.26\,\mbox{GHz}$.
                  In the limits of the roton azimuthal mode, the frequencies differ from
       $\nu_{rot}$ by $\pm 1.2\,\mbox{MHz}$, whereas the frequencies $\nu_{l_{rot}}$ for a resonator in He-II and in vacuum
       differ by $\sim 20\,\mbox{MHz}$ \cite{d}. Therefore, the exact condition of resonance must give
       $\nu_{l_{rot}} = \nu_{rot} \pm 1\,\mbox{MHz}$. It is easy to prove that condition (\ref{gal})
        is sufficiently close to the exact one.

          Consider the field $\textbf{A}$ in helium.  Near the disk, it satisfies Eq. (\ref{2b}),
           whose solution at $A_{z}=0$ looks as
                 \begin{eqnarray}
  &&\textbf{A}_{h}(\rho ,\varphi,z,t) =  \sum\limits_{l}\int dQ_{z}g_{l}(Q_{z})F(Q_{z},z)
   e^{il\varphi -i\omega t} \nonumber \\
   &\times &\left \{[J_{l-1}(Q^{h}_{\rho}\rho)+a_{l}(Q_{z})N_{l-1} (Q^{h}_{\rho}\rho)](\textbf{e}_{\rho}+i\textbf{e}_{\varphi})
    \right.  \nonumber \\    &+&   \left.
    [\tilde{c}_{l}(Q_{z})J_{l+1} (Q^{h}_{\rho}\rho) +b_{l}(Q_{z})N_{l+1} (Q^{h}_{\rho}\rho)]\right. \label{4-h} \\
    &\times & \left. (-\textbf{e}_{\rho} +i\textbf{e}_{\varphi})
   \right \}+ \mbox{c.c.}
     \nonumber \end{eqnarray}
            To determine $\textbf{A}_{h},$
            we use solution (\ref{Ain1}),
            (\ref{Ain2}) for the field inside the disk and the BCs
            (\ref{gr1}) and (\ref{gr2}).

             a) Regions above and under the disk, $|z|>h_{d}/2, \rho < R_d$.
              Here, we neglect the Neumann functions in (\ref{4-h}), because they increase unboundedly
                        as $\rho \rightarrow 0.$
              As solutions for $F(Q_{z},z),$ we can take functions of the form $e^{i\alpha z}$ or $e^{\alpha
              z}$.
                           Relations (\ref{gr1}) and (\ref{gr2}) imply that
               the solutions $\textbf{A}$ on the disk surface must coincide for the disk
               and helium, $\textbf{A}_h =
               \textbf{A}_d$. Therefore, relation (\ref{4-h}) is reduced to the form
               \begin{eqnarray}
  \textbf{A}_h &=&  \frac{A_{m}}{8}
  e^{i(l\varphi - \omega t)}e^{-\kappa_{z}(|z|-h_{d}/2)} \nonumber \\
   &\times & \left \{[J_{l-1} (Q^{h}_{1}\rho)-c_{l}J_{l+1} (Q^{h}_{1}\rho)]\textbf{e}_{\rho}
     \right.  \label{9}  \\ &+& \left.
   i\textbf{e}_{\varphi}[J_{l-1} (Q^{h}_{1}\rho)+c_{l}J_{l+1} (Q^{h}_{1}\rho)]\right \}+ \mbox{c.c.}
        \nonumber \end{eqnarray}
              The sewing $\textbf{A}_h = \textbf{A}_d$ on
         the disk surface requires that $Q^{h}_{1}$ coincide with
         $Q_{1}$ from (\ref{Ain2}).  For (\ref{9}), we have
         $Q^{h}_{1}=\sqrt{\frac{\omega^2}{c^2}\varepsilon_{h} +
           \kappa_{z}^2}$, and the condition $Q^{h}_{1}=Q_{1}$
           gives $ \kappa_{z}\approx 6.53/h_d$.

     b) Region in helium around the disk, $|z|\leq h_{d}/2, \rho \geq
             R_d$. Relations (\ref{gr1}) and (\ref{gr2}) yield $\textbf{A}_{\varphi}^h =
             \textbf{A}_{\varphi}^d$ and $\textbf{A}_{\rho}^h =
             \textbf{A}_{\rho}^d =0$.
              Then only the harmonic with $Q_z = \pi/\tilde{h}_d$ and $l=66$ remains in (\ref{4-h}) in the sum $\sum\limits_{l}\int dQ_{z},$ and
             the function $F(Q_{z},z)$ is reduced to $\cos{(z\pi/\tilde{h}_d)}.$ In this case, we have for the roton frequency:
              \begin{equation}
  Q^{h}_{1} =  \sqrt{(Q^{h})^2- Q_{z}^2}\approx 24.626/R_{d},
       \label{12a} \end{equation}
    \begin{equation}
     Q^{h} = \omega/c_{h}, \quad Q_{z} =\pi/\tilde{h}_{d}.     \label{12b} \end{equation}
           At such $Q^{h}_{1},$
    values of the Neumann functions (see Fig. 1) in (\ref{4-h}) are greater by ~20 orders than
    values of the Bessel functions for the region with helium near the disk. Therefore, the latter must be neglected (by the physical reasoning, solution (\ref{4-h}) should be written in terms of the Hankel functions; since
    the Bessel functions are small in them, only the Neumann functions remain). As a result, we obtain
    \begin{eqnarray}
  \textbf{A}_h &\approx &  g\cdot A_{m}e^{il\varphi -i\omega t}
       \cos{(\pi z/\tilde{h}_d)} \nonumber \\
       &\times & \left \{[N_{l-1}(Q^{h}_{1}\rho)-b_{l}N_{l+1} (Q^{h}_{1}\rho)]\textbf{e}_{\rho}
      \right.  \label{11}  \\ &+& \left.
   i\textbf{e}_{\varphi}[N_{l-1} (Q^{h}_{1}\rho)+b_{l}N_{l+1} (Q^{h}_{1}\rho)]\right \}+ \mbox{c.c.}
         \nonumber \end{eqnarray}
      The functions $N_{l}(x)$ have the asymptotics \cite{abr}
         \begin{equation}
    N_{l}(x\rightarrow 0)\approx -\frac{(l-1)!}{\pi}\left (\frac{2}{x} \right )^l \equiv N^{as}_{l}(x).
         \label{n-as} \end{equation}
         The numerical analysis indicates that, for values of the argument $x\sim Q^{h}_{1}R_{d}\approx 24.6,$
            this asymptotics is approximately (with a correction coefficient) satisfied, namely:
         $N_{67}(Q^{h}_{1}R_{d})\approx 10.38N^{as}_{67}(Q^{h}_{1}R_{d})\approx -1.59\cdot 10^{20}$,
         $N_{67}(Q^{h}_{1}R_{d}+1\mbox{mm})\approx 17.65N^{as}_{67}(Q^{h}_{1}R_{d}+1\mbox{mm})\approx -3.3\cdot 10^{17}$,
         $N_{65}(Q^{h}_{1}R_{d})\approx 11.2N^{as}_{65}(Q^{h}_{1}R_{d})\approx -6.05\cdot 10^{18}$,
         $N_{65}(Q^{h}_{1}R_{d}+1\mbox{mm})\approx 19.39N^{as}_{65}(Q^{h}_{1}R_{d}+1\mbox{mm})\approx -1.57\cdot 10^{16}$.
         The condition $\textbf{A}_{\rho}^h =
             \textbf{A}_{\rho}^d =0$ is satisfied at $b_{l}=N_{l-1}(Q^{h}_{1}R_{d})/N_{l+1} (Q^{h}_{1}R_{d})$,
            which gives  $b_{66}\approx 0.0381$.
         We can avoid great numbers in solution (\ref{11}), if it is rewritten in the normalized form and by taking
         the condition $\textbf{A}_{\varphi}^h = \textbf{A}_{\varphi}^d$ into account:
          \begin{eqnarray}
  \textbf{A}_h & \approx &  A_{m}J_{l-1}(Q_{1}R_{d})e^{il\varphi -i\omega t}
       \cos{(\pi z/\tilde{h}_d)} \nonumber \\
       & \times & \left \{ [n_{l-1}(Q^{h}_{1}\rho)-b_{l}n_{l+1} (Q^{h}_{1}\rho)]\textbf{e}_{\rho}
      \right. \label{ah2a}  \\ &+& \left.
   i\textbf{e}_{\varphi}[n_{l-1} (Q^{h}_{1}\rho)+b_{l}n_{l+1} (Q^{h}_{1}\rho)]\right \}+ \mbox{c.c.},
         \nonumber \end{eqnarray}
         where $l=66$, $J_{l-1}(Q_{1}R_{d})\approx 1/27.831$, and
           \begin{equation}
  n_{l\pm 1}(Q^{h}_{1}\rho) =  N_{l\pm 1}(Q^{h}_{1}\rho)/N_{l-1}(Q^{h}_{1}R_{d}).
         \label{n-n} \end{equation}
                 For the region with helium, relation (\ref{ah2a}) can be approximately written
          near the disk in a simple form
          \begin{eqnarray}
  \textbf{A}_h & \approx & i\textbf{e}_{\varphi}A_{m}2J_{l-1}(Q_{1}R_{d})e^{il\varphi -i\omega t}
      \nonumber \\ &\times & \cos{(\pi z/\tilde{h}_d )} (R_{d}/\rho)^{l-1}+ \mbox{c.c.}
         \label{ah2b} \end{eqnarray}

 c)  In the region $|z|\geq h_{d}/2, \rho \geq
               R,$ we sew together the solutions for the regions $|z|>h_{d}/2, \rho <
               R$ and $|z|\leq h_{d}/2, \rho \geq R$ along the surface of their intersection.
               This surface is symmetric relative to a turn around the $z$ axis and intersects any of the planes
                $z, \rho$ along a certain curve $z(\rho)$ which cannot be calculated analytically.
                Moreover, the analysis indicates that the intersection happens not for all $z$ and $\rho$. This means
                 that the solution is more complicated in this transient region and cannot be determined by the separation of variables.
                Below, we will use a rough sewing, by considering that there exists a line $z(\rho)$, along which a smooth sewing is realized.
                Such an approximation is apparently admissible, because the field is small in this region.
                             The final solution for the field in helium near the disk has the form
                \begin{equation}
  \textbf{A}_h = \textbf{A}^{h}_{0}+ \mbox{c.c.},
         \label{13} \end{equation}
          \begin{eqnarray}
  \textbf{A}^{h}_{0} &\approx & A_{m} e^{i(l\varphi - \omega t)} \left [a_{1}(\rho,z)(\textbf{e}_{\rho}+
  i\textbf{e}_{\varphi}) \right. \nonumber \\
  &+& \left. a_{2}(\rho,z)(-\textbf{e}_{\rho}+
  i\textbf{e}_{\varphi})\right ],
         \label{13b} \end{eqnarray}
   \begin{equation}
  a_{1}(\rho,z) \approx
\left [ \begin{array}{ccc}
    \frac{1}{8} e^{-\kappa_{z}(|z|-h_{d}/2)} J_{l-1} (Q_{1}\rho)   & \   (I),  & \\
    \frac{1}{27.831}\cos{(z\pi/\tilde{h}_d)}n_{l-1}(Q^{h}_{1}\rho)    & \ (II), &
   \label{14a} \end{array} \right. \end{equation}
   \begin{equation}
 a_{2}(\rho,z) \approx
\left [ \begin{array}{ccc}
    \frac{c_{l}}{8} e^{-\kappa_{z}(|z|-h_{d}/2)} J_{l+1} (Q_{1}\rho)  & \   (I),  & \\
    \frac{b_{l}}{27.831}\cos{(z\pi/\tilde{h}_d)}n_{l+1} (Q^{h}_{1}\rho)  & \ (II), &
   \label{14b} \end{array} \right. \end{equation}
   where $I, II$ stand for the regions ($I: |z|\geq h_{d}/2, \rho =0 \div \rho(z)$; $II:  \rho \geq R_{d}, |z|= 0 \div |z(\rho)|$), and $z(\rho)$
   or $\rho(z)$ is the sewing line. In this case,
   $l=66$, $c_{66}\approx 0.265$, $b_{66}\approx 0.0381$, $\tilde{h}_d \approx 1.087 h_{d}$,
   $\kappa_{z}\approx 6.53/h_{d}$,  $1/27.831 = J_{l-1}(Q_{1}R_{d})$,  $Q_{1}\approx 72.143/R_{d}$,
        $Q^{h}_{1}\approx 24.626/R_{d}$,   $A_{m}=2.94 E_{m}c/\omega$, $E_{m}\simeq
           10^4\,\mbox{V}/\mbox{m}$ (value of $E_{m}$ in the International System of units is taken from the experiment \cite{d} for the frequency band of a generator
           ${ \sss\triangle} \nu_{pump} \simeq 50\,\mbox{kHz}$).

      The presented solution is in an approximate agreement with experiment.
      Only one difference can be noticed: according to the experiment, the field $\textbf{A}_h$ decreases by 1--2 orders, as the distance from the disk
      increases by 1 mm. From (\ref{13})--(\ref{14b}), we obtain that the attenuation in regions I and II is as high as $\sim 700$ and $\sim 43$ times, respectively.
      That is, the attenuation is too strong in region I.
           However, since we used the approximate experimental data on the field,
        the divergence can be related just to this circumstance.

                 In practice, each resonance mode is a very narrow band of frequencies, for which the EM field has shape of a
         ``dome''. Most likely, this testifies to the resonance amplification of modes with some
         dispersion of $Q_z$ and $Q_{\rho}$  (independently). Respectively, the resonance frequency is eroded with the formation of a dome. But solution
         (\ref{13})--(\ref{14b}) does not consider the dome and implies that the EM field has a single frequency,
        so that the coefficients are found for the frequency $\nu_{rot}=180.3\,\mbox{GHz}$.

         According to quantum electrodynamics \cite{berest}, in order to
        quantize the electromagnetic field, it is necessary to know the photon wave functions (WF) $\Psi^n_{phot}$ ($n$ is a collection of
        quantum numbers characterizing a state of a photon) which form the basis, in which the general
        solution of the Maxwell equations for a specific physical system is expanded.
         The collection of basis functions depends on the symmetry of the system.
         Therefore, photons are of different types --- plane,
         circular (or cylindrical), or spherical  --- and are characterized by different collections of quantum numbers.
         If the system is translationally invariant, then it is convenient to expand the field $\textbf{A}$
        in plane waves. In this case, a photon has a definite momentum $\hbar\textbf{Q}$ and a definite energy $\hbar\omega$, and
         $Q=\omega/c$.
        In the case under consideration, the disk violates the translational symmetry. However, the circular symmetry holds.
         Respectively, a solution of the Maxwell equations for $\textbf{A}$ takes form (\ref{Ain1}),
            (\ref{Ain2}), (\ref{13})--(\ref{14b}). Whence we determine the WF of a photon with $Q_{z}=\pi/\tilde{h}_d, n_{\rho}=1$
            for the $\rho $- and $\varphi $-polarizations:
           \begin{equation}
         \Psi_{phot}^{\rho,\varphi}  = e^{il\varphi -i\omega t}
                  \cos{(z\pi/\tilde{h}_d)}
         \left [J_{l-1} (Q_{1}\rho) \mp c_{l}J_{l+1}
         (Q_{1}\rho)\right ]
             \label{vf-ph1}  \end{equation}
            inside the disk, and
         \begin{equation}
         \Psi_{phot}^{\rho,\varphi}  = e^{il\varphi -i\omega t}
          (a_{1}(\rho,z)\mp a_{2}(\rho,z))
             \label{vf-ph2}  \end{equation}
        ouside the disk,   where the upper sign in the parentheses is related to the $\rho$-polarization (normalizing factors are omitted).

    With regard for the angular momentum operator
      $\hat{\textbf{L}} = -i\hbar [\textbf{r}\times\nabla]$  (in
      particular, $\hat{L_z} = -i\hbar\partial/\partial\varphi$), it is easy to prove that the WF $\Psi_{phot}$
      is characterized by eigenvalues $E=\hbar\omega$ and
                $L_z =\hbar l$. However, the momentum for states (\ref{vf-ph1}), (\ref{vf-ph2}) is not defined.
           We call such states ``circular photons'' (c-photons).

         Hence, a resonator creates some number of identical c-photons with  $l=66$,
          $n_{\rho}=1,$ and $Q_{z}=\pi/\tilde{h}_d$. In space, a c-photon is localized in the disk and near it.
         We note that such a photon \textit{cannot be represented} as a
         superposition of plane photons. Indeed, let
          the EM field $\textbf{A}_h$ in helium be expanded in plane waves
         with the wave vector $k=c_{h}/\omega$. Since the disk and helium have different values of $\varepsilon,$
         a photon, being plane in helium, is reflected from the
         cylindrical surface of the disk in the form of a fan of diverging
         almost radial waves and
         is refracted in a complicated way inward the disk as a system of
         converging waves. Therefore, photons are not plane in helium or in the disk.
        As follows from the laws of conservation,
         a quasiparticle created by a c-photon in helium must have
         the same quantum numbers as the c-photon
         (energy and angular momentum), but it has no momentum
         in the disk plane. This implies that a phonon
         created by a c-photon in helium must also possess the circular symmetry.
          In this case, a c-photon \textit{emitted} by a resonator can be approximately represented
          as a collection of radially moving almost plane photons,
          the last being wave packets with size $\sim \lambda_{phot}$.
          Such photons can create plane rotons or phonons,
          if the disk or, as was assumed in \cite{svh3}, helium absorbs a recoil momentum. But this is already a
          combined process,
         and its probability is much less than that of the
         direct c-photon $\rightarrow $ c-phonon process.

          \section{Circular phonons}
          In helium far from the disk, ordinary ``plane'' phonons and rotons, being
          wave packets localized in space, are propagating. But, near the disk and also far from it in the case where
          $\lambda$ of a phonon is of the order of the disk size, the structure of a phonon must correspond to the symmetry of the disk.

           As an example, we consider a free particle in the space
        with an infinite cylinder with radius $R_{c}$. We assume that the particle cannot penetrate into
        the cylinder. Therefore, its WF $\Psi(\textbf{r},t)$ satisfies the BC
       \begin{equation}
       \Psi(\rho =R_{c},t)=0
        \label{c0} \end{equation}
       and the Schr\"{o}dinger equation
    \begin{equation}
       i\hbar\frac{\partial \Psi}{\partial t}= - \frac{\hbar^2}{2m}\triangle \Psi.
     \label{c3} \end{equation}
      In the stationary case where $\tilde{\Psi}(\textbf{r})  = e^{i\omega t}\Psi (\textbf{r},t),$ the Schr\"{o}dinger equation
      takes the form of a wave equation
       \begin{equation}
       \triangle \tilde{\Psi} + k^2 \tilde{\Psi}= 0, \quad  k^2 =
       2m\omega/\hbar.
     \label{c4} \end{equation}
     Solutions of Eqs. (\ref{c0}) and (\ref{c4}) are the functions
       \begin{equation}
  \tilde{\Psi}(l,k_z,k_{\rho}) =  e^{-ik_{z}z}e^{il\varphi}[
  a_{l}H^{(1)}_{l}(k_{\rho}\rho)+b_{l}H^{(2)}_{l}(k_{\rho}\rho)],
           \label{c5} \end{equation}
       where $k_{\rho}^2 = k^2 - k_z^2$, $H^{(1)}_{l}(x)=J_{l}(x)+iN_{l}(x)$ and
       $H^{(2)}_{l}(x)=J_{l}(x)-iN_{l}(x)$ are the Hankel functions, and $a_{l}$ and $b_{l}$ are selected so that
       $a_{l}H^{(1)}_{l}(k_{\rho}R_{c})+b_{l}H^{(2)}_{l}(k_{\rho}R_{c})=0$.
       In view of the asymptotics
     $H^{(1)}_{l}(x\rightarrow \infty)=\sqrt{2/\pi x}\cdot exp(ix-i\pi l/2-i\pi/4)$ and
     $H^{(2)}_{l}(x\rightarrow \infty)=\sqrt{2/\pi x}\cdot exp(-ix+i\pi l/2+i\pi/4),$ these functions describe
      the diverging and converging waves, respectively.
     Thus, if an impenetrable cylinder is present in space, the solution for a
     \textit{free} particle
     is represented by circular waves (\ref{c5}) (with various  $l$ and $k_{z}$), rather than plane ones.
     The solution differs from a plane wave, because the interaction
      is indirectly present through BC.

      If a disk is present instead of a cylinder, and
      $\Psi=0$ on its whole surface, then the solutions of Eq. (\ref{c4}) take the other
      form:
       \begin{equation}
  \tilde{\Psi}(l,k_z,k_{\rho}) =  \left (e^{-ik_{z}z}+e^{ik_{z}z}\right
  )e^{il\varphi}J_{l}(k_{\rho}\rho),
            \label{c6} \end{equation}
       \begin{equation}
   k_{\rho}^2 = k^2 - k_z^2,  \ k_{z}=\left (\frac{1}{2}+n_{z}\right )\frac{2\pi}{h_{d}}, \ n_{z}=0,\pm 1,\pm 2,...,
   \label{c8} \end{equation}
  and $J_{l}(k_{\rho}R_{d})=0$. Like that in Section 2, the Neumann function $N_{l}$ is not present in the solution, since it
   increases  unboundedly  near the disk axis.

   We now consider helium surrounding the disk. The microscopic
      model for He-II without disk is constructed in the main
   (see, e.g., survey \cite{obz}) for
    periodic BCs, as the volume of the system tends to infinity. The model involves the WFs of the ground state $\Psi_{0}$ and a state with one phonon $\Psi_{c}\Psi_{0}$.
     A specific feature of our problem consists in the presence of a disk in helium.
                It would be most proper to find the functions $\Psi_{0}$ and $\Psi_{c}\Psi_{0}$ with zero BCs realized in the nature
        and with regard for a disk. This requires to construct the full microscopic model of He-II with a disk, which is a very complicated problem.
          Therefore, we limit ourselves by the calculation of $\Psi_{c}$
        for an infinite system without regard for BCs. But, in this case, it will be necessary in certain situations to pass from $\int d\textbf{k}$ to
              the sum $\sum\limits_{\textbf{k}}$ and to know the value of $\Psi_{c}\Psi_{0}$ on boundaries.
             To his end, we will consider that, according to the preliminary analysis, the zero BCs lead to the equations
               \begin{equation}
   \sin{(k_{z}|z|+\alpha)}|_{z=\pm h_{d}/2, \pm H/2}=0,
   \label{bc1} \end{equation}
   \begin{equation}
   J_{l}(k_{\rho}R_{d})=J_{l}(k_{\rho}R_{\infty})=0,
   \label{bc2} \end{equation}
        which yield the conditions of quantization for $k_{\rho}$ and $k_{z}$:
         \begin{equation}
   k_{z}=\frac{2\pi n_{z}}{H-h_{d}}, \ n_{z}=\pm 1,\pm 2,...,
   \label{bc3} \end{equation}
     \begin{equation}
       k_{\rho}=\frac{\pi n_{\rho}}{R_{\infty}-\tilde{R}_{d}}, \ n_{\rho}\gg l
   \label{bc4} \end{equation}
   ($n_{z}\neq 0$, since $\Psi_{c}\Psi_{0}$ will not be zero on the $z$-boundaries otherwise).
   Here, $R_{\infty}$ is the radius of a container with helium, and $\tilde{R}_{d}$  depends on $n_{\rho}$: for the least $n_{\rho}=1,$
   relation (\ref{bc5}) yields $\tilde{R}_{d}\approx 1.5R_{d},$ and $\tilde{R}_{d}$ decreases to $R_{d}$ with increase in $n_{\rho}$.
    At small $n_{\rho}$ ($\lsim l$), there exists no solution $k_{\rho}$, for which the relations
    $J_{l}(k_{\rho}R_{d})=0$ and  $J_{l}(k_{\rho}R_{\infty})=0$ from (\ref{bc2}) would be simultaneously satisfied.
    However, the symmetry of the system deviates from the cylindrical one near the container walls. Therefore, the relation
    $J_{l}(k_{\rho}R_{\infty})=0$ should not apparently hold, and only $J_{l}(k_{\rho}R_{d})=0$ is valid.
     This yields
      \begin{equation}
       k_{\rho}(l,n_{\rho})= \mu_{l}^{(n_{\rho})}/R_{d},  \  n_{\rho}=1,2,3,... \lsim l.
      \label{bc5} \end{equation}
  The last relation can be rewritten in the form of (\ref{bc4}), by introducing $\tilde{R}_{d}$.
    Though conditions (\ref{bc1})--(\ref{bc5}) will be used, we will find the WF of a phonon
    in a simpler approximation, by neglecting the BCs (as usually the  micromodels of He-II are constracted \cite{obz}).

     It follows from the $N$-particle Schr\"{o}dinger equation that if
    the WF of the ground state of He-II is represented in the form
     $\Psi_{0}=const\cdot e^{S}$, then the WF $\Psi_{\textbf{k}}$ of a
    plane (p-) or circular (c-) phonon satisfies the equation
    \begin{equation}
   - \frac{\hbar^2}{2m_{4}}\sum\limits_{j=1}^N [\nabla_{j}^{2}+2(\nabla_{j}S)\nabla_{j}]
      \Psi_{\textbf{k}}=E(\textbf{k})\Psi_{\textbf{k}}.
     \label{c10} \end{equation}
     We now consider the zero approximation without
     any interaction between atoms. In this case,
     $S$=const, and (\ref{c10}) is reduced to
      \begin{equation}
   - \frac{\hbar^2}{2m_{4}}\sum\limits_{j=1}^N \nabla_{j}^{2}\Psi_{\textbf{k}}=E(\textbf{k})\Psi_{\textbf{k}}
     \label{c11} \end{equation}
   which is the Schr\"{o}dinger equation for $N$ free particles. For a
   translationally invariant system in the case where a single particle has a momentum $\hbar \textbf{k}$, and
   the rest ones are immovable, a solution of the equation looks as
    \begin{equation}
     \Psi_{\textbf{k}} = \rho_{-\textbf{k}}=\frac{1}{\sqrt{N}}\sum\limits_{j=1}^N
     e^{i\textbf{k}\textbf{r}_{j}}, \quad
     E_{0}(\textbf{k})=\frac{\hbar^{2}\textbf{k}^{2}}{2m_{4}},
     \label{c12} \end{equation}
      where $\rho_{\textbf{k}}$ are the ``plane'' collective variables,
     and $m_{4}$ is the helium atom mass.  This solution serves as the zero approximation for the WF of a p-phonon.
     The consideration of the interaction leads, as known, to the transformation of the one-particle excitations (\ref{c12})
     to collective ones: $\Psi_{\textbf{k}}$ acquires corrections nonlinear in $\rho_{\textbf{k}},$
     and $E_{0}(\textbf{k})\sim k^2$ is replaced by a more complicated dispersion law for quasiparticles. If a disk is present in helium, the
     translational symmetry is broken, but the circular symmetry holds. Therefore, according to (\ref{c4}) and
     (\ref{c6}), the solution of (\ref{c11}) is the WF
      \begin{eqnarray}
  \Psi_{c}(l,k_z,k_{\rho}) & =&
  \frac{c_{l,k_z,k_{\rho}}}{\sqrt{N}} \sum\limits_{j=1}^N  e^{i(l\varphi_{j}-\omega t)}e^{ik_{z}z_{j}}
  J_{l}(k_{\rho}\rho) \nonumber \\
  &\equiv& \sum\limits_{j=1}^{N}f_{l,k_{z},k_{\rho}}(\textbf{r}_{j}) \equiv \rho_{c}(l,k_z,k_{\rho}).
            \label{c9} \end{eqnarray}
     It is the zero approximation to the WF of a cir\-cular  phonon in helium-II with the immersed disk
     (the summation is made over all atoms). We omit the Neumann function, since namely function (\ref{c9}) is a solution
      under the most correct zero BCs.
      The consideration of the interaction between atoms will lead to the appearance of corrections to (\ref{c9}) which are nonlinear in $\rho_{c},$ but we will neglect them.
       In (\ref{c9}), $\rho_{c}$ means the circular
     collective variables.

    The following question is of importance: Does the energy $E_{c}(\textbf{k})$ of a c-phonon coincide with
    the energy $E(\textbf{k})$ of a p-phonon at the same $k$?
          For a free particle (Eq. (\ref{c4})), the energy
        depends only on  $k$ (but not on $k_z$ and $k_{\rho}$ separately).
        Analogously, the energy of a c-phonon must depend only on
        $k$. But, at $k_{\rho}\ll k,$ a c-phonon is close to a
          p-phonon, and, hence, its energy must be close to
          the energy of a p-phonon, by differing proportionally to  smallest
           $k_{\rho}/k$.  Therefore, for any other $k_{\rho},$
          the energy of a c-phonon must also  be close to
         the energy $E(\textbf{k})$ of a p-phonon. Generally speaking, the exact
         equality $E_{c}(k)=E(k)$ is possible as well.

    Acting by the momentum operator $\hat{L_z} =
    -i\hbar\sum\limits_{j}\partial/\partial\varphi_{j}$ on the WF
    of a c-phonon (\ref{c9}), we verify that a c-phonon possesses the intrinsic momentum
    $\hat{L_z} = \hbar l$.
      Thus, in what follows, we will use the
      zero approximation (\ref{c9}) and the conditions of quantization (\ref{bc3})--(\ref{bc5}) for the WF of a c-phonon.

            \section{Normalization of the wave function of a circular phonon}
       As seen from
        (\ref{c9}), we need to know the coefficient  $c_{l,k_z,k_{\rho}}$ (below, $\tilde{c}$) for the WF of a c-phonon. We will determine it from the condition of normalization
          \begin{equation}
           \int d\Omega |\Psi_{c}\Psi_{0}|^2 =   1,
         \label{n1} \end{equation}
     where $d\Omega = d\textbf{r}_{1}\ldots d\textbf{r}_{N}$, and $\textbf{r}_{j}$ are coordinates of the $j$-th atom.
             Using (\ref{c9}), we have
             \begin{equation}
              \int d\Omega |\Psi_{c}\Psi_{0}|^2 = I_{1}+I_{2},
     \label{n2} \end{equation}
         \begin{equation}
             I_{1}= \frac{\tilde{c}^2}{N}\int d\Omega \Psi_{0}^{2}\sum\limits_{j=1}^{N}J_{l}^{2}(k_{\rho}\rho_{j})  =
             \tilde{c}^2 \int d\Omega \Psi_{0}^{2}J_{l}^{2}(k_{\rho}\rho_{1}),
     \label{n3} \end{equation}
        \begin{eqnarray}
             I_{2} &=& \frac{\tilde{c}^2}{N}\int d\Omega \Psi_{0}^{2}2\sum\limits_{j_{1}<j_{2}}e^{il(\varphi_{j_{1}}-\varphi_{j_{2}})}
             e^{ik_{z}(z_{j_{1}}-z_{j_{2}})}\nonumber \\
             &\times & J_{l}(k_{\rho}\rho_{j_{1}})J_{l}(k_{\rho}\rho_{j_{2}}).
     \label{n4} \end{eqnarray}
         In the real experiment, the disk is positioned in helium between two long cylindrical rods-antennas
          located in the disk plane at a distance of $1.37R_{d}$ from the disk center.
          Near the antennas, a
       phonon wave loses the circular shape. But the analytic calculation of a new shape is a quite difficult problem, and we will neglect
       the difference of the symmetry of the system from the cylindrical one, by considering that the container with helium is a cylinder with radius
       $R_{\infty}$ and height $H$ (then the helium volume $V=\pi R_{\infty}^{2}H$).

          As known, the pair correlation function  $g(\textbf{r}_{1},\textbf{r}_{2})$ determining the probability
          for atom 1 to be at a point $\textbf{r}_{1}$ and for atom 2 to be at a point $\textbf{r}_{2}$
          is presented by the integral
             \begin{equation}
              \int d\textbf{r}_{3}\ldots d\textbf{r}_{N} \Psi_{0}^2 = g(\textbf{r}_{1},\textbf{r}_{2})/V^2.
     \label{n10a} \end{equation}
       For a translationally invariant system,
       $g(\textbf{r}_{1},\textbf{r}_{2}) =       g(\textbf{r}_{1}-\textbf{r}_{2})$.
       In our case, a disk positioned in He-II breaks the translational invariance. But, at
       small $|\textbf{r}_{1}-\textbf{r}_{2}|,$ the function
       $g(\textbf{r}_{1},\textbf{r}_{2})$  is determined by the interaction of the nearest atoms, so that
       it should depend in helium with a disk only
       on the difference
       $\textbf{r}_{1}-\textbf{r}_{2}$, if $\textbf{r}_{1}$ and
       $\textbf{r}_{2}$ are not too close to the disk. For a region
       near the disk (at distances of about several interatomic ones), $g(\textbf{r}_{1},\textbf{r}_{2})\neq
       g(\textbf{r}_{1}-\textbf{r}_{2})$. But it is a very thin layer which hardly influences the processes in bulk.
       Therefore, we accept that the relation $g(\textbf{r}_{1},\textbf{r}_{2})=
       g(\textbf{r}_{1}-\textbf{r}_{2})$ is always true, and, hence,
       \begin{equation}
              \int d\textbf{r}_{3}\ldots d\textbf{r}_{N} \Psi_{0}^2 = g(|\textbf{r}_{1}-\textbf{r}_{2}|)/V^2.
     \label{n10} \end{equation}
           For atoms which are not located at the disk surface, the relation
       \begin{equation}
          S(k) = 1+\frac{N}{V} \int [g(r)-1]e^{-i\textbf{k}\textbf{r}}d\textbf{r}
          \label{n11} \end{equation}
            obtained for translationally invariant systems is also valid.
          In addition, if $\textbf{r}_{1}$ is far from the disk, then the relation
       \begin{equation}
              \int d\textbf{r}_{2}\ldots d\textbf{r}_{N} \Psi_{0}^2 = \int d\textbf{r}_{2}g(\textbf{r}_{1},\textbf{r}_{2}) = 1/V
     \label{n5} \end{equation}
         is true. Indeed, this integral determines the probability to find atom 1 at the point
         $\textbf{r}_{1}$, and it is obvious that the probability for all points far from the disk is the same.
                     With regard for (\ref{n5}), we obtain that integral (\ref{n3}) is
           \begin{eqnarray}
             I_{1} &=& \frac{\tilde{c}^2}{V} \int d\textbf{r}_{1}J_{l}^{2}(k_{\rho}\rho_{1}) = \frac{2\tilde{c}^2}{R_{\infty}^2}
             \int\limits_{0}^{R_{\infty}}\rho d\rho J_{l}^{2}(k_{\rho}\rho) \nonumber \\
             &\approx &  \frac{2\tilde{c}^2}{\pi k_{\rho} R_{\infty}}B(k_{\rho}R_{\infty},l),
     \label{n6} \end{eqnarray}
      where
           \begin{equation}
      B(x,l) = \pi x \int\limits_{0}^{1}ydy J_{l}^{2}(y\cdot x).
                \label{aF1}  \end{equation}

              We note that the Bessel functions satisfy the relation \cite{tih}
    \begin{eqnarray}
          & &  \int \limits_{0}^{R}\rho d\rho J_{l}\left (\frac{\mu_{l}^{(m_{1})}\rho}{R}\right )
             J_{l}\left (\frac{\mu_{l}^{(m_{2})}\rho}{R}\right )  \nonumber \\
             &=&  \delta_{m1,m2}\frac{R^2}{2}\left [J_{l}^{\prime}(\mu_{l}^{(m_{1})}) \right ]^2,
     \label{n7} \end{eqnarray}
     where $\delta_{m1,m2}$ is the Kronecker symbol, and  $J_{l}^{\prime}(x)= \frac{d}{dx}J_{l}(x)$.
     First, we consider large $k_{\rho}$.  At $k_{\rho}\rho \gg l,$
            the condition (\ref{bc4})  is valid and the following asymptotic is  true \cite{y}:
          \begin{equation}
   J_{l}(k_{\rho}\rho) \approx \sqrt{\frac{2}{\pi k_{\rho}\rho}}\left [\cos{\alpha^{l}(k_{\rho}\rho)}
   -\frac{4l^{2}-1}{8k_{\rho}\rho}\sin{\alpha^{l}(k_{\rho}\rho)}\right ],
      \label{c14} \end{equation}
     \begin{equation}
    \alpha^{l}(k_{\rho}\rho)= k_{\rho}\rho -\frac{\pi l}{2}-\frac{\pi}{4}.
   \label{c15} \end{equation}
          In this case,
       the function $J_{l}(k_{\rho}\rho)$ performs
       many oscillations on the interval $\rho =0\div R_{\infty}$, and the following relation more general than (\ref{n7}) is approximately valid:
           \begin{equation}
              \int\limits_{0}^{R_{\infty}}\rho d\rho J_{l}(k_{\rho}\rho)J_{l}(q_{\rho}\rho) \approx
              \delta_{k_{\rho},q_{\rho}}\frac{R_{\infty}}{\pi k_{\rho}} B(k_{\rho}R_{\infty},l).
                   \label{n15-2} \end{equation}
       Moreover, at $k_{\rho}R_{\infty}\gg l,$ we have
              \begin{equation}
      B(k_{\rho}R_{\infty},l) \approx \frac{\pi k_{\rho} \tilde{R}_{\infty}}{2}
             \left [J_{l}^{\prime}(k_{\rho}\tilde{R}_{\infty}) \right ]^2,
                \label{aF2}  \end{equation}
     where $\tilde{R}_{\infty}$ is a value of $\rho$ which is the closest to $R_{\infty}$ and is such that
     $k_{\rho}\tilde{R}_{\infty}$ is equal to one of the zeros $\mu_{l}^{(j)}$ of the Bessel function $J_{l}(x)$.
     According to (\ref{bc2}),  $\tilde{R}_{\infty}=R_{\infty}$.  Relation (\ref{c14}) yields
         \begin{equation}
   J_{l}^{\prime}(k_{\rho}\rho) \approx -\sqrt{\frac{2}{\pi k_{\rho}\rho}}\sin{\left [\alpha^{l}(k_{\rho}\rho)\right ]}
    + O\left ((k_{\rho}\rho)^{-3/2} \right ).
      \label{n8} \end{equation}
              For
       $\rho=R_{\infty},$ we have $J_{l}(k_{\rho}\rho)=0$;
       therefore, $\cos{\alpha^{l}(k_{\rho}R_{\infty})}\approx 0$, which yields
        \begin{equation}
        [J_{l}^{\prime}(k_{\rho}\tilde{R}_{\infty}) ]^{2}\mid _{k_{\rho}R_{\infty}\gg l} \approx
       2/\pi k_{\rho}R_{\infty},
        \label{aF3} \end{equation}
         \begin{equation}
         B(k_{\rho}R_{\infty},l)\mid _{k_{\rho}R_{\infty}\gg l} \approx  1,
        \label{aF4} \end{equation}
       \begin{equation}
       I_{1}\mid _{k_{\rho}R_{\infty}\gg l} \approx  \frac{2\tilde{c}^2}{\pi k_{\rho}R_{\infty}}.
          \label{n9} \end{equation}
          At small $k_{\rho}$  ($k_{\rho}R_{\infty}\lsim l$), (\ref{bc5}) is satisfied.
          In this case, the function  $B(k_{\rho}R_{\infty},l)$ can be determined only numerically. The analysis indicates that $k_{\rho}$ from (\ref{bc5})
            satisfies the relation $B(k_{\rho}R_{\infty},l=66) \gsim 0.9$. In other words, for all $k_{\rho},$ we may take
            $B(k_{\rho}R_{\infty},l=66) \approx 1$.

                     With regard for (\ref{n10}) and the well-known relations
           \begin{equation}
          J_{l}(x) =\frac{i^l}{2\pi} \int\limits_{-\pi+\beta}^{\pi+\beta}
          e^{-ix\cos{\psi}\pm il\psi}d\psi,
          \label{n12} \end{equation}
           \begin{equation}
          \delta(x) =\frac{1}{2\pi} \int\limits_{-\infty}^{\infty}
          e^{-ikx}dk,
          \label{n13} \end{equation}
          \begin{equation}
          g(r) = 1+ \frac{V}{N(2\pi)^3} \int (S(q)-1)e^{i\textbf{q}\textbf{r}}d\textbf{q}
          \label{n14} \end{equation}
          ($\beta$ is arbitrary; (\ref{n14}) follows from (\ref{n11})), let us write the vector $\textbf{q}$ in the CCS.
          Then integral (\ref{n4}) reads
           \begin{eqnarray}
             I_{2} &=& \frac{\tilde{c}^2}{\pi R_{\infty}^2}\int d\varphi_{q}q_{\rho}dq_{\rho}\rho_{1}d\rho_{1}\rho_{2}d\rho_{2}
             [S(q_{\rho},-k_{z})-1] \nonumber \\
             &\times & J_{l}(k_{\rho}\rho_{1})J_{l}(k_{\rho}\rho_{2})J_{l}(q_{\rho}\rho_{1})J_{l}(q_{\rho}\rho_{2}).
           \label{n15} \end{eqnarray}
             Using conditions (\ref{bc4}) and (\ref{bc5}) for $q_{\rho}$,  we pass from $\int dq_{\rho}$ to the
              sum $\frac{\pi}{R_{\infty}-\tilde{R}_{d}}\sum\limits_{q_{\rho}}$.
                         In this case,
              $q_{\rho}$ and $k_{\rho}$ in (\ref{n15}) are quantized identically.
              With regard for (\ref{n15-2}), we finally get
                                     \begin{equation}
             I_{2} \approx
             \frac{2\tilde{c}^{2}(S(k)-1)B^{2}(k_{\rho}R_{\infty},l)}{\pi k_{\rho} (R_{\infty}-\tilde{R}_{d})}.
           \label{n16} \end{equation}
             By using relations (\ref{n1}), (\ref{n2}), (\ref{n6}), and (\ref{n16}),
              we obtain the required result for the normalization of the WF of a c-phonon:
             \begin{equation}
             \tilde{c} \equiv c_{l,k_z,k_{\rho}} \approx \sqrt{\frac{0.5\pi k_{\rho}
             R_{\infty}}{B+B^{2}(S(k)-1)(1-\tilde{R}_{d}/R_{\infty})^{-1}}},
           \label{n18} \end{equation}
           where $B\equiv B(k_{\rho}R_{\infty},l)$.
    This formula is approximately true also for arbitrary $k_{\rho}$ which is not quantized according to
    (\ref{bc4}) and (\ref{bc5}). We note that, while integrating, we do not consider that helium atoms cannot be present in the
           volume occupied by the disk, but taking this circumstance into account does not practically change the integrals and result (\ref{n18}).

             \section{Conclusion}
             Thus, we have determined the distributions of the electromagnetic field inside and outside a resonator,
             as well as the wave function of a circular  phonon. Without these quantities, it is impossible to calculate the SHF absorption spectrum of liquid helium
             which arises due to the creation of quasiparticles in helium by the field of a resonator. In our opinion, just the mutual transformation of excitations
              with the circular symmetry (photon $\Leftrightarrow $ phonon or photon $\Leftrightarrow $ roton) allows one to understand the
               process of absorption in helium with an immersed disk resonator, when the momentum conservation law is formally broken, and it is necessary to determine
              which quantum numbers of created and disappeared quasiparticles must be conserved. The calculation of the probabilities of relevant
              transitions and the description of the phenomena discovered in experimental works \cite{svh2,svh3} are planned to present in the subsequent publications.

       \vskip3mm
        The authors are grateful to
        V.\,N. Derkach, E.\,Ya. Rudavskii,  A.\,S. Rybalko,
      and Yu.\,V. Shtanov for the useful discussions.

     \renewcommand\refname{}

\end{document}